\documentclass[a4paper,11pt]{article}
\usepackage[british]{babel}
\usepackage[utf8]{inputenc}
\usepackage{pos}
\usepackage{subcaption}
\usepackage{siunitx}
\usepackage{enumitem}
\usepackage{braket}

\title{BMW/DMZ calculation of the hadronic vacuum polarisation for the muon magnetic moment}

\author*[a,b]{F.M.~Stokes}
\author[b,c]{A.~Cotellucci}
\author[d]{M.~Davier}
\author[e,c,f,b,g,h]{Z.~Fodor}
\author[c]{F.~Frech}
\author[b,i]{D.~Giusti}
\author[b]{A.Yu.~Kotov}
\author[j]{L.~Lellouch}
\author[k]{B.~Malaescu}
\author[j,l]{S.~Mutzel}
\author[b,c]{K.K.~Szabo}
\author[b,c]{B.C.~Toth}
\author[j]{G.~Wang}
\author[d]{Z.~Zhang}
\onbehalf{for the BMW and DMZ collaborations}

\affiliation[a]{Special Research Centre for the Subatomic Structure of Matter, Department of Physics,\\
  Adelaide University / Tirkangkaku, Kaurna Country, South Australia 5005, Australia}
\affiliation[b]{J\"ulich Supercomputing Centre, Forschungszentrum J\"ulich,
D-52428 J\"ulich, Germany}
\affiliation[c]{Department of Physics, University of Wuppertal, D-42119 Wuppertal, Germany}
\affiliation[d]{IJCLab, Universit\'e Paris-Saclay et CNRS/IN2P3, Orsay, 91405, France}
\affiliation[e]{Physics Department, Pennsylvania State University, University Park, PA 16802, USA}
\affiliation[f]{Institute for Computational and Data Sciences, Pennsylvania State University,\\
University Park, PA 16802, USA}
\affiliation[g]{Institute for Theoretical Physics, E\"otv\"os University, H-1117 Budapest, Hungary}
\affiliation[h]{University of California, San Diego, 9500 Gilman Drive, La Jolla, CA 92093, USA}
\affiliation[i]{Fakult\"at f\"ur Physik, Universit\"at Regensburg, 93040, Regensburg, Germany}
\affiliation[j]{Aix Marseille Univ, Universit\'e de Toulon, CNRS, CPT, IPhU, Marseille, France}
\affiliation[k]{LPNHE, Sorbonne Universit\'e, Universit\'e Paris Cit\'e, CNRS/IN2P3, Paris, 75252, France}
\affiliation[l]{Laboratoire de Physique de l’Ecole Normale Sup\'erieure, Mines Paris - PSL,\\
CNRS, Inria, PSL Research University, Paris, France}

\emailAdd{f.stokes@adelaide.edu.au}

\abstract{For twenty years, a persistent discrepancy between experimental
measurements and theoretical calculations of the muon anomalous magnetic
moment have provided tantalising hints of new physics. In recent years,
improvements to the experimental precision have appeared to make the
tension stronger and stronger. However, at the same time, our lattice calculation
overturned the theoretical consensus, completely eliminating the tension.
I will present the latest results from the Budapest-Marseille-Wuppertal (BMW)
and DMZ collaborations, with a determination of the hadronic vacuum polarisation
contribution to a precision of 0.45\%}

\FullConference{The 42nd International Symposium on Lattice Field Theory (LATTICE2025)\\
2-8 November 2025\\
Tata Institute of Fundamental Research, Mumbai, India\\}

\newcommand{\lsim}{ {\
\lower-1.2pt\vbox{\hbox{\rlap{$<$}\lower5pt\vbox{\hbox{$\sim$}}}}\ } }
\newcommand{\gsim}{ {\
\lower-1.2pt\vbox{\hbox{\rlap{$>$}\lower5pt\vbox{\hbox{$\sim$}}}}\ } }

\def\amulohvp{a_\mu^\text{LO-HVP}}

\def\mev{\mathrm{Me\kern-0.1em V}}
\def\gev{\mathrm{Ge\kern-0.1em V}}
\def\tev{\mathrm{Te\kern-0.1em V}}

\def\mev{\mathrm{Me\kern-0.1em V}}
\def\gev{\mathrm{Ge\kern-0.1em V}}
\def\tev{\mathrm{Te\kern-0.1em V}}

\def\refcite#1{Ref.~\cite{#1}}

\def\reff#1{\ref{#1}}

\def\fig#1{Fig.~\reff{#1}}

\def\sec#1{Section~\reff{#1}}
\def\tab#1{Table~\reff{#1}}

\def\amulohvp{a_{\mu}^\text{LO-HVP}}

\begin{document}
\maketitle

\section{Overview}
\begin{figure}[b!]
    \centering
    \includegraphics[width=\textwidth]{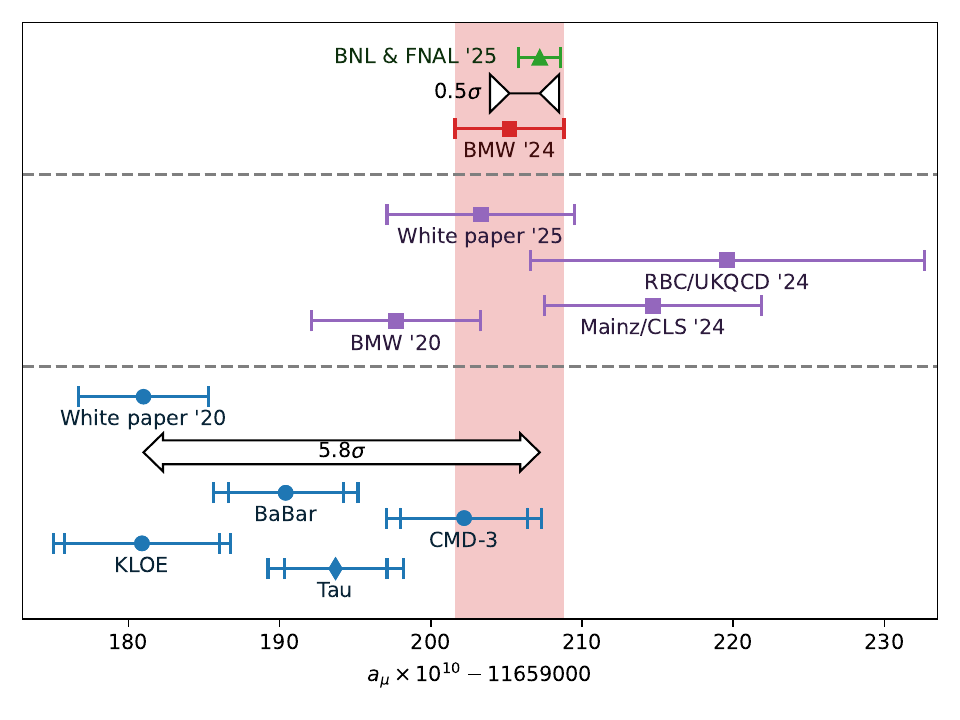}
    \vspace{-1em}
    \caption
    {
	Comparison of Standard Model predictions for the muon anomalous magnetic
	moment with its measured value, taken from \refcite{Boccaletti:2024guq}.
	The top panel shows a comparison of the world-average experimental
	measurement of $a_\mu$~\cite{Muong-2:2025xyk} with the Standard Model
	prediction obtained by the BMW collaboration~\cite{Boccaletti:2024guq},
	denoted by the red band.
	The middle panel shows a predictions based on the earlier BMW result from
	\refcite{Borsanyi:2020mff} as well as results from two other similar
	calculations, from RBC/UKQCD~\cite{RBC:2018dos,RBC:2023pvn,RBC:2024fic}
	and Mainz~\cite{Djukanovic:2024cmq}, along with the
	consensus combination of the three from \refcite{Aliberti:2025beg}.
	The lower panel shows the predictions for $\amulohvp$ obtained in the
	2020 approach~\cite{Aoyama:2020ynm} using specific experimental
	inputs~\cite{Davier:2023fpl}. These correspond to
	BaBar~\cite{BaBar:2009wpw,BaBar:2012bdw},
	KLOE~\cite{KLOE:2008fmq,KLOE:2010qei,KLOE:2012anl,KLOE-2:2017fda} and
	CMD-3~\cite{CMD-3:2023alj}, and $\tau$
	decays~\cite{Davier:2009ag,Davier:2013sfa}.
	Note, all Standard Model predictions include non-HVP contributions from
	``White paper '25'', except for ``White paper '20''.
    }
    \label{fig:amucomp}
\end{figure}

Almost twenty years ago, physicists at Brookhaven National Laboratory
measured the magnetic moment of the muon with a remarkable precision of
0.54 parts per million (ppm)~\cite{Muong-2:2006rrc}.
Since that time, the reference Standard Model prediction for this
quantity has exhibited a persistent discrepancy with experiment of more
than three sigma~\cite{Aoyama:2020ynm}.
This raises the tantalising possibility of undiscovered forces or
elementary particles.

The attention of the world was drawn to this discrepancy when
Fermilab presented a brilliant confirmation of Brookhaven's measurement,
which brought the discrepancy to 4.2 sigma~\cite{Muong-2:2021ojo}.
In the meantime a very large-scale lattice QCD calculation of a key
theoretical contribution was performed by the
Budapest-Marseille-Wuppertal (BMW) collaboration~\cite{Borsanyi:2020mff}, as
seen in \fig{fig:amucomp}.
This result significantly reduces the difference between theory and
experiment, suggesting that new physics may not be needed to
explain the experimental results.
However, it simultaneously introduces a new discrepancy
with the existing data-driven determination of this contribution.

Since then, the experimental~\cite{Muong-2:2025xyk} results have been
updated with significantly improved precision, and the lattice result
has been independently confirmed by other lattice collaborations.
At the same time, new developments in the data-driven inputs that the
lattice calculations replace~\cite{CMD-3:2023alj,Davier:2009ag,
Davier:2013sfa} lead to a significant spread in the results depending
on what inputs are taken.
This has culminated in an updated theory prediction based on the lattice
results instead of the data-driven determinations, as seen in
\fig{fig:amucomp}.

In these proceedings, I present a new hybrid calculation that
combines an update to the most precise lattice results with data-driven
inputs in a low-energy region where the observed discrepancies are not
present. This new result leads to a prediction that differs from the
experimental measurement by only $0.5\sigma$, providing a remarkable
validation of the Standard Model to 0.31 ppm.

\section{Muon magnetic moment}
The quantity under investigation is the so-called anomalous magnetic
moment, which is defined in terms of the magnetic moment
\[
    \vec{\mu}_\mu = g_\mu \left(\frac{q}{2m}\right) \vec{S}\,.
\]
where $q$ is the charge, $m$ is the mass, $\vec{S}$ is the spin, and $g$ is the so-called $g$ factor.
The anomalous magnetic moment is the relative deviation of $g$ from its tree-level value of 2.
\[
    a_\mu = \frac{g_\mu-2}{2}\,.
\]

This anomalous magnetic moment is measured experimentally by injecting a
polarised beam of positively-charged muons into a large magnetic storage
ring and observing the decay of the muons into positrons.
As the muons cycle around the ring, their magnetic moments precess.
The difference between the precession and orbital frequencies is directly
related to the anomalous magnetic moment.
It can be observed as a variation in the energy distribution of the
positrons as the momentum and magnetic moment come in and out of
alignment.
This is how the experimental result (green triangle) in
\fig{fig:amucomp} was obtained.

The theoretical prediction is dominated by contributions from
electromagnetism, which have been computed in perturbation theory up to
five loops~\cite{Aoyama:2019ryr,Aoyama:2012wk}.
The much smaller electroweak contributions have been computed to two
loops~\cite{Czarnecki:2002nt,Gnendiger:2013pva}.
These provide extremely precise values for these parts of the total, as
can be seen in \tab{tab:contributions}.

\begin{table}[tbp]
    \centering
    \begin{tabular}{rS[table-format=8.4]cS[table-format=1.4]}
	Contribution & \multicolumn{3}{c}{Value} \\
	\hline\noalign{\vskip2pt}
        \(a_{\mu}^\mathrm{QED} \times 10^{10}\) & 11 658 471.88 & \(\pm\) & 0.02 \\\noalign{\vskip-0.2pt}
        \(a_{\mu}^\mathrm{EW} \times 10^{10}\) & 15.44 & \(\pm\) & 0.04 \\\noalign{\vskip-0.2pt}
        \(a_{\mu}^\mathrm{HVP} \times 10^{10}\) & 704.5 & \(\pm\) & 6.1 \\\noalign{\vskip-0.2pt}
        \(a_{\mu}^\mathrm{HLbL} \times 10^{10}\) & 11.55 & \(\pm\) & 0.99 \\\noalign{\vskip2pt}
	\hline\noalign{\vskip1pt}
	\(a_{\mu} \times 10^{10}\) & 11 659 203.3 & \(\pm\) & 6.2 \\\noalign{\vskip1pt}
	\hline
    \end{tabular}

    \caption
    {
	Individual contributions to the Standard Model theory consensus from \refcite{Aliberti:2025beg}.
    }
    \label{tab:contributions}
    \vspace{-1em}
\end{table}

The remaining contributions come from the strong force, and are: the
hadronic vacuum polarisation (HVP), and the hadronic light by light
scattering (HLbL).
These effects contribute a very small amount to the total value, but
they dominate the uncertainty, as seen in \tab{tab:contributions}.
The work presented in these proceedings computed the HVP, which has the
largest contribution to the uncertainty, and is where there have been
significant theoretical developments.

\section{Hadronic Vacuum Polarisation}\label{sec:hvp}
Until recently, the accepted approach to obtain the HVP was a data-driven
evaluation, based on experimental measurements of hadron production at
electron-positron colliders~\cite{Aoyama:2020ynm,Davier:2017zfy,
Keshavarzi:2018mgv,Colangelo:2018mtw,Hoferichter:2019mqg}.
These experimental results were related to the HVP using a dispersive
approach.

\begin{figure}[bp!]
    \centering
    \includegraphics[width=\textwidth]{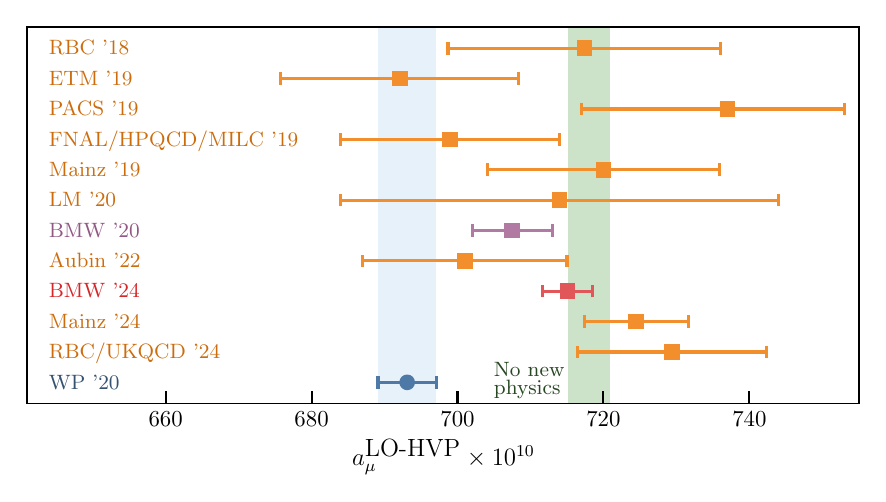}
    \vspace{-2.3em}
    \caption
    {
	Comparison of lattice values for the leading HVP contribution to the muon magnetic moment.
	The blue circle is the Standard Model consensus from the 2020 White Paper~\cite{Aoyama:2020ynm},
	and the squares are lattice results. The older lattice results~\cite{Budapest-Marseille-Wuppertal:2017okr,
	RBC:2018dos,Giusti:2019hkz,Shintani:2019wai,FermilabLattice:2019ugu,Gerardin:2019rua,
	Lehner:2020crt,Aubin:2022hgm}
	are consistent with both the Standard Model consensus and the experimental measurement.
	The four most precise results (from BMW~\cite{Borsanyi:2020mff,Boccaletti:2024guq},
        Mainz~\cite{Djukanovic:2024cmq}, and RBC/UKQCD~\cite{RBC:2024fic}) agree with the experimental
        measurement but not the	Standard Model consensus.
    }
    \label{fig:latticecomp}
\end{figure}

The HVP can also be calculated in lattice QCD from the vector-vector
correlator using the time-momentum representation~\cite{Bernecker:2011gh}.
It is computed as an integral over Euclidean time,
\begin{align}
    C(t) &= \frac{1}{3} \sum_{\mu=1,2,3} \braket{J_{\mu}(t) J_{\mu}(0)}\\
    a_{\mu}^\mathrm{LO-HVP} &= \alpha^2 \int_0^\infty K(t) C(t) dt
\end{align}
where \(K(t)\) is a kernel function that encodes the kinematics of the
muon.

Historically, lattice QCD results had uncertainties significantly larger
than the data-driven results.
These uncertainties were large enough that the results were generally
consistent with both the data-driven results and the experimental
measurement, and hence were unable to weigh in on the discrepancy.
This can be seen in \fig{fig:latticecomp}.

In 2020, we completed a lattice calculation of the HVP with sub-percent
precision~\cite{Borsanyi:2020mff}.
This result was the first lattice calculation with errors comparable to
the data-driven evaluations, and was the first that was precise enough
to resolve the discrepancy between theory and experiment.
To our surprise, this result was significantly larger than the data-driven
evaluation, and was close to the experimental measurement.

In 2024, we completed a new calculation~\cite{Boccaletti:2024guq}
that significantly improved upon our 2020 precision, surpassing the
data-driven evaluations.
Independent calculations from Mainz~\cite{Djukanovic:2024cmq}, and
RBC/UKQCD~\cite{RBC:2024fic} confirmed these results, albeit with
larger uncertainties.

Our 2020 result was a $3.4\times$ increase in precision compared
to our previous work~\cite{Budapest-Marseille-Wuppertal:2017okr},
and the 2024 update was a further $1.6\times$ improvement.
Many upgrades were required to reach this level of precision.
These improvements were possible by extensive work from many
lattice groups around the world in developing and refining the
techniques required.

\begin{figure}[tb]
    \centering
    \includegraphics[scale=0.9]{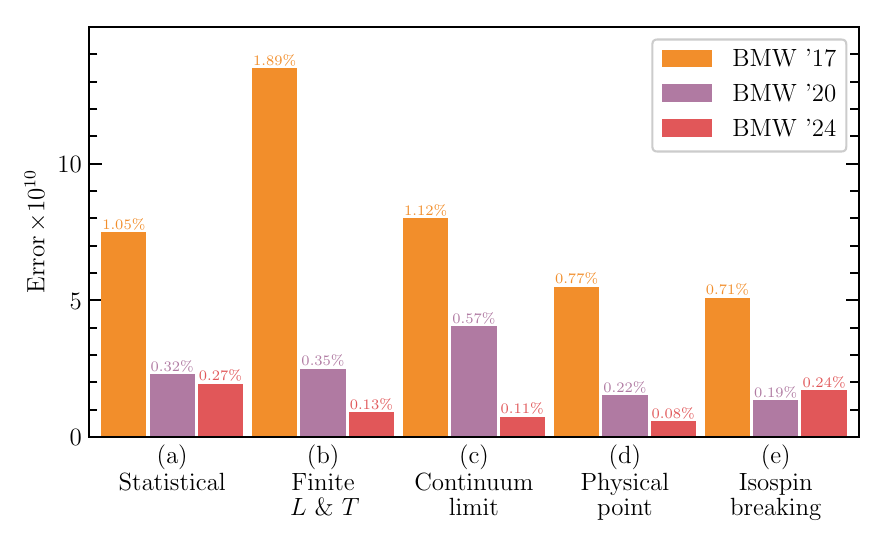}
    \caption
    {
	Main uncertainties and their reduction in the BMW collaboration's successive lattice
	calculations of $\amulohvp$.  Their sources are labelled (a-e) in the
	text and are given a short descriptive title below the bars in the
	plot. Their approximate size relative to the total LO-HVP contribution
	obtained in the present work is also shown. The orange bars on the left
	of each group correspond to the 2017 BMW result~\cite{Borsanyi:2017zdw},
	the purple ones in the middle to the 2020 findings~\cite{Borsanyi:2020mff},
	and the red ones on the right to the latest BMW preprint~\cite{Boccaletti:2024guq}.
    }
    \label{fig:err_improve}
\end{figure}

Figure~\ref{fig:err_improve} summarises the five main sources of
uncertainty in our three calculations, and how they were improved
across the three publications to reach the current level of precision.
These sources of uncertainty are:
\pagebreak
\begin{enumerate}[label=\alph*)]
    \itemsep0em
    \item the statistical uncertainty;
    \item the effects of the finite volume and temporal extent of the
	lattice;
    \item uncertainties in the continuum extrapolation;
    \item uncertainties on the physical values for the inputs used to
	set the quark masses and scale; and
    \item isospin-breaking effects from QED and the up-down quark mass
	difference.
\end{enumerate}

The dominant uncertainties in the 2017 work came from source (b):
uncertainties in removing the effect of the \qty{6}{fm} box in which we perform
our simulations.
In 2020, these effects were addressed using dedicated large-volume
simulations in an \qty{11}{fm} box.
Our finite-size effects are exponentially suppressed, so these
large-volume simulations see dramatically smaller finite-volume effects.
The remaining uncertainties due to residual finite volume effects at
\qty{11}{fm}, and to uncertainties in the \qty{11}{fm} simulation results were
small, and mostly affected the large-Euclidean-time tail of the integrand.

Once this uncertainty was reduced, the other sources became significant.
In order to bring the final uncertainty down further, we now needed to
tackle them.
Four further improvements in the 2020 work allowed these reductions:
\begin{enumerate}
    \item Algorithmic improvements~\cite{Bali:2009hu,Blum:2012uh}
	allowed increased statistics, improving the statistical and
	continuum limit uncertainties.
    \item New techniques for the exact treatment of infrared
	modes~\cite{Neff:2001zr} improved statistical uncertainty.
    \item A new scale-setting input (the $\Omega$ baryon mass)
	reduced the uncertainty of the physical point.
    \item Calculations of all isospin-breaking contributions
	significantly reduced the associated uncertainty.
\end{enumerate}

In 2024, the analysis was updated with new simulations at a finer
lattice spacing, and an overhaul to the analysis procedure.
In this work, the Euclidean-time integral was broken up into
different windows, and the continuum limit was performed for each
window independently.
This allowed for a separation of the dominant uncertainties, with
different Euclidean times being affected by different sources of
uncertainty:
\begin{enumerate}
    \item The short-distance part has complicated discretisation effects
	that make the continuum limit challenging, but the statistical
	errors are small, so it is easy to constrain the form of this
	extrapolation.
    \item The long-distance part has large statistical and finite-size
	uncertainties.
    \item Most of the total value comes from intermediate distances with
	small statistical errors, and simple well-constrained continuum
	limits.
\end{enumerate}

Separating out the different distance scales like this not only improved
the uncertainties in our lattice calculations, but also allowed us to
take a novel ``hybrid'' approach that brings together the strengths of
the lattice and data-driven approach in a complementary way. This approach
will be the focus of \sec{sec:hybrid} of these proceedings.

Our 2024 calculation was performed completely blind, with all HVP
data multiplied by unknown random numbers before analysis commenced.
These blinding factors were only removed once the analysis was finalised
and the manuscript was almost complete.
This ensured the result was free from any conscious or unconscious biases
from any of the researchers.
As seen in \fig{fig:latticecomp}, this result is in good agreement
with the recent lattice results from Mainz~\cite{Djukanovic:2024cmq} and
RBC/UKQCD~\cite{RBC:2024fic}, the latest white paper theory
consensus~\cite{Aliberti:2025beg} and with the experimental
measurement~\cite{Muong-2:2025xyk}.

\section{Lattice Setup}

\begin{figure}
    \centering
    \includegraphics{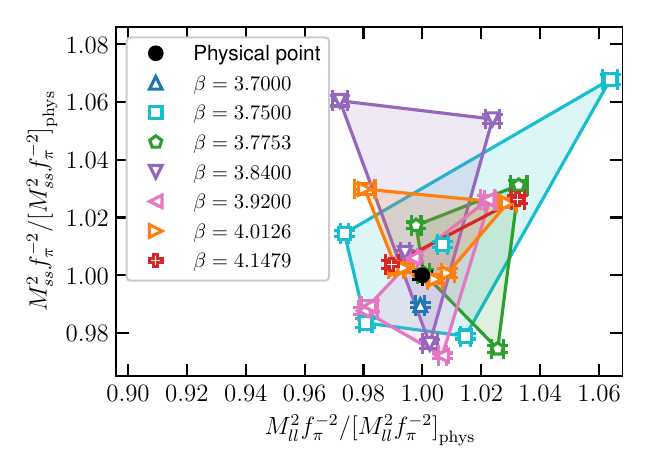}
    \caption
    {
	\label{fig:landscape} Spread of our ensembles around the physical
	point, as defined by the masses of the pseudo-scalar mesons \(M_{ll}\)
	and \(M_{ss}\).
	Different colours denote different lattice spacings.
	The black point denotes the (isospin-symmetric) physical point, with
	error bars corresponding to the uncertainties from our determination
	of \(M_{ll}\) and \(M_{ss}\) in physical units.
    }
\end{figure}

The simulations for all three works were performed using the tree-level
Symanzik gauge action~\cite{Luscher:1984xn} and a one-link staggered
fermion action for \(2+1+1\) dynamical quark flavours.
Where the gauge link appears in the fermion action, we apply four steps
of stout smearing~\cite{Morningstar:2003gk} with a smearing parameter of
\(\rho = 0.125\).

\begin{table}[tbp]
    \centering
    \begin{tabular}{c|c|c|c|c|c}
	$\beta$&$a$ [fm]&$L/a \times T/a $& $am_s$&$m_s/m_l$&\#confs\\
	\hline\hline
	3.7000&0.1315&$48\times 64$&0.057291&27.899&904\\
	\hline
	3.7500&0.1191&$56\times 96$&0.049593&28.038&315\\
	&&&0.049593&26.939&516\\
	&&&0.051617&29.183&504\\
	&&&0.051617&28.038&522\\
	&&&0.055666&28.083&215\\
	\hline
	3.7753&0.1116&$56\times 84$& 0.047615&27.843&510\\
	&&&0.048567&28.400&505\\
	&&&0.046186&26.469&507\\
	&&&0.049520&27.852&385\\
	\hline
	3.8400&0.0952&$64\times 96$&0.043194&28.500&510\\
	&&&0.043194&30.205&436\\
	&&&0.040750&28.007&1503\\
	&&&0.039130&26.893&500\\
	\hline
	3.9200&0.0787&$80\times 128$& 0.032440&27.679&506\\
	&&&0.034240&27.502&512\\
	&&&0.032000&26.512&1001\\
	&&&0.032440&27.679&327\\
	&&&0.033286&27.738&1450\\
	&&&0.034240&27.502&500\\
	\hline
	4.0126&0.0640&$96\times 144$&0.026500&27.634&446\\
	&&&0.026500&27.124&551\\
	&&&0.026500&27.634&2248\\
	&&&0.026500&27.124&1000\\
	&&&0.027318&27.263&985\\
	&&&0.027318&28.695&1750\\
	\hline
	4.1479&0.0483&$128\times 192$&0.019370&27.630&2792\\
	&&&0.019951&27.104&2225\\
    \end{tabular}
    \caption
    {
	\label{tab:configs} List of the ensembles used in our 2024 work, with gauge
	coupling, lattice spacing, lattice size, strange-quark
	mass, ratio of strange and light quark masses and number of
	configurations.
    }
\end{table}

As shown in \fig{fig:landscape}, the mass of the light (\(m_u=m_d=m_l\))
and strange (\(m_s\)) quarks are chosen to scatter about the physical point.
The charm mass parameter is set by the ratio $m_c/m_s = 11.85$, from
\refcite{McNeile:2010ji}.
We perform simulations at seven different lattice spacings as listed in
\tab{tab:configs}.
The finest lattice spacing was added for the 2024 calculation.

\begin{table}[tbp]
    \centering
    \begin{tabular}{c|c|c|c|c|c}
	$\beta$&$a$ [fm]&$L/a \times T/a $& $am_s$&$m_s/m_l$&\#confs\\
	\hline\hline
	3.7000&0.1315&$24\times 48$&0.057291&27.899&716\\
	&&$48\times 64$&0.057291&27.899&300\\
	\hline
	3.7753&0.1116&$28\times 56$& 0.047615&27.843&887\\
	\hline
	3.8400&0.0952&$32\times 64$&0.043194&28.500&1110\\
	&&&0.043194&30.205&1072\\
	&&&0.040750&28.007&1036\\
	&&&0.039130&26.893&1035\\
    \end{tabular}
    \caption
    {
	\label{tab:qedconfigs} List of the ensembles used in our evaluation
	of dynamical QED effects, with gauge coupling, lattice spacing,
	lattice size, strange-quark mass, ratio of strange and light quark
	masses and number of configurations.
    }
\end{table}

Our isospin symmetric simulations all have a spatial extent of approximately
\qty{6}{fm} and a temporal extent of approximately \qty{9}{fm}.
For computing dynamical QED effects, we use a different set of ensembles,
some of which have smaller volumes, as listed in \tab{tab:qedconfigs}.

\begin{table}[tbp]
    \centering
    \begin{tabular}{c|c|c|c|c|c}
	$\beta$&$a$ [fm] & $am_s$&$m_s/m_l$&$L/a \times T/a$&\#confs\\
	\hline\hline
	0.7300&0.112&0.060610&44.971&$56\times 84$&7709\\
	&&&&$96\times 96$&962\\
	&&&33.728&$56\times 84$&8173\\
	&&&&$56\times 84$&813\\
    \end{tabular}
    \caption
    {
	\label{tab:4hexconfigs} List of ensembles used for our large-volume
	study in 2020 and 2024, with gauge coupling, lattice spacing,
	lattice size, strange-quark mass, ratio of strange and light quark
	masses and number of configurations.
    }
\end{table}

To address finite volume effects, we have dedicated large-volume simulations
with a spatial extent of approximately \qty{11}{fm}.
These ensembles are computed using the DBW2 gauge action~\cite{Takaishi:1996xj}
and a one-link staggered fermion action for \(2+1\) dynamical quark flavours,
with four steps of HEX smearing~\cite{DeGrand:2002vu} applied to the gauge links.
This formulation drastically reduces the ultraviolet fluctuations in the gauge
configurations, significantly reducing taste-breaking effects in the staggered
quarks. These simulations are performed at a single lattice spacing, as listed
in \tab{tab:4hexconfigs}.

\section{Analysis Strategy}

To obtain the HVP contribution in the continuum physical point, we perform
global fits to its lattice spacing and mass dependence. We consider a variety
of fit forms and use the variation between the different fit forms as a
systematic uncertainty.

We consider lattice spacing dependence of the form
\begin{equation}
    A_0 + A_2 a^2 \alpha_s^\gamma(a) + A_4 a^4 + A_6 a^6
\end{equation}
where we get different fit variations by \(A_6\) or \(A_6\) and \(A_4\) to
zero, and by choosing different values of \(\gamma\) in the set
\({0.0, 0.5, 1.0, 1.5, 2.0, 2.5}\). We consider further variations by
excluding some of the coarsest lattice spacings from the fit. When we do
this, we always keep at least one more lattice spacing than fit parameters.

We also take into account the quark mass dependence by considering terms
linear in the variables
\begin{equation}
    X_l = M_{ll}^2 w_0^2 - [M_{ll}^2 w_0^2]_\mathrm{phys} \qquad and \qquad X_s = M_{ss}^2 w_0^2 - [M_{ss}^2 w_0^2]_\mathrm{phys}
\end{equation}
which describe the deviation from the physical light and strange quark mass
respectively. Finally, we make a number of variations to the input into the
fits, for example varying the physical inputs within their uncertainties.

Once we have all of these fits, how do we combine them to get a value and
uncertainty? We need to take into account the quality of the fits as well
as the number of parameters in order to avoid over- or under-fitting. To do
this, we use a weight based on a version of the Akaike information criterion
(AIC)~\cite{Akaike1973,Akaike:1974vps,Akaike1978c} with modifications
derived in \refcite{Borsanyi:2020mff} to account for differing numbers of
data points in different fits:
\begin{equation}
    w= \exp\left[-\frac{1}{2}\left(\chi^2 + 2n_\mathrm{par}-n_\mathrm{data}\right)\right]\label{eq:aic}\ .
\end{equation}
This weight is computed from the \(\chi^2\) of the fit, the number of parameters
\(n_{\mathrm{par}}\), and the number of data points included in the fit
\(n_{\mathrm{data}}\).

Using jackknife resampling, we obtain the statistical uncertainties on each fit
and combine the corresponding statistical distributions with their AIC weights
to form a total CDF. Then we obtain a central value from the median of this CDF
and a total uncertainty (combined statistical and systematic) from the width of
this distribution. Specifically, we look at the one (and two) sigma confidence
band, and take half (a quarter) of its width as the uncertainty. To avoid any
non-gaussianity of this distribution leading to underestimated uncertainties,
we take the larger of these two uncertainty estimates.

\section{Euclidean-Time Windows}

\begin{figure}[b!]
    \centering
    \includegraphics[scale=0.8]{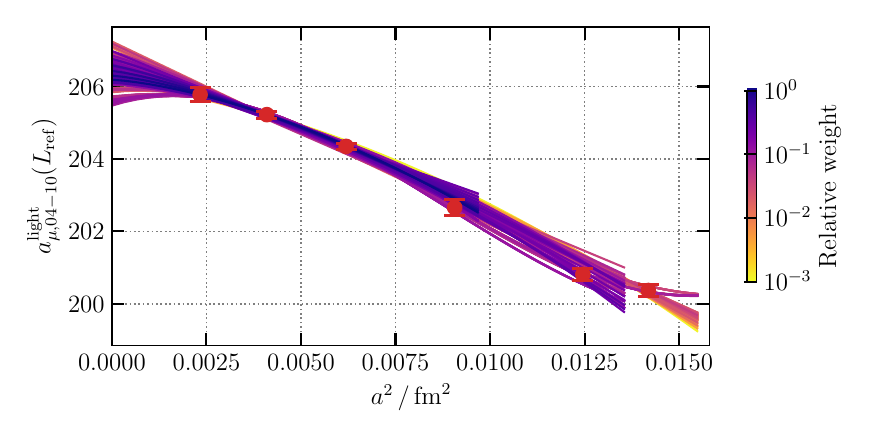}\\
    \includegraphics[scale=0.8]{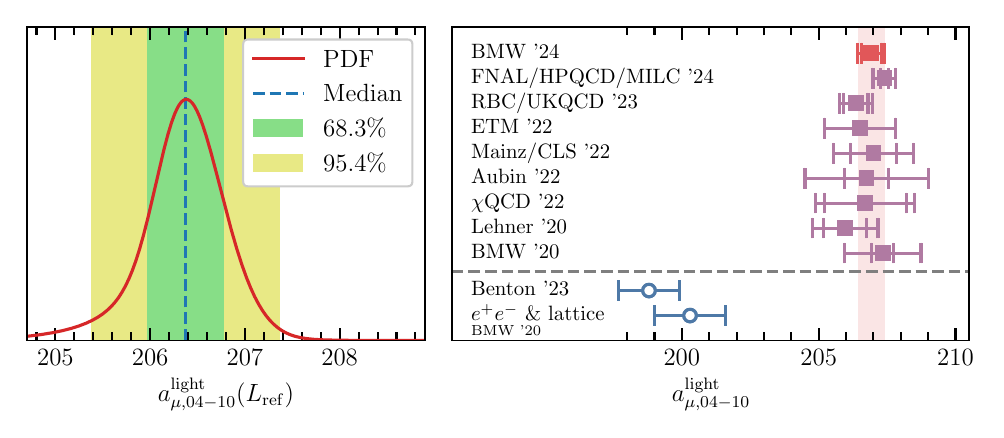}
    \caption
    {
	\label{fig:idwin_light} Light-connected
	window observable $a_{\mu,04-10}^\mathrm{light}$.
	Top: continuum extrapolations, coloured according to the
	weight of the fit in our analysis.
	Lower left: the probability distribution
	function, which shows a single well-defined peak.
	The shaded bands correspond to the one- and two-sigma confidence bands.
	Lower right: we compare our result with others from the literature, both
	lattice~\cite{Borsanyi:2020mff,Lehner:2020crt,
	Wang:2022lkq,Aubin:2022hgm,Ce:2022kxy,
	ExtendedTwistedMass:2022jpw,RBC:2023pvn,MILC:2024ryz} and
	data-driven~\cite{Borsanyi:2020mff,Benton:2023dci}.
    }
\end{figure}

As discussed in \sec{sec:hvp}, the various challenges facing this
calculation affect different Euclidean time ranges differently. We can take
advantage of this by breaking the integral up into different Euclidean time
windows~\cite{RBC:2018dos}. We do this by modifying the kernel function
\(K(t)\) to focus on specific regions in \(t\). It is straightforward to
compute the effect of such a modification upon the data-driven calculation,
so calculations of particular windows can be compared to equivalent
data-driven results.

\begin{figure}[tbp]
    \centering
    \includegraphics[scale=0.8]{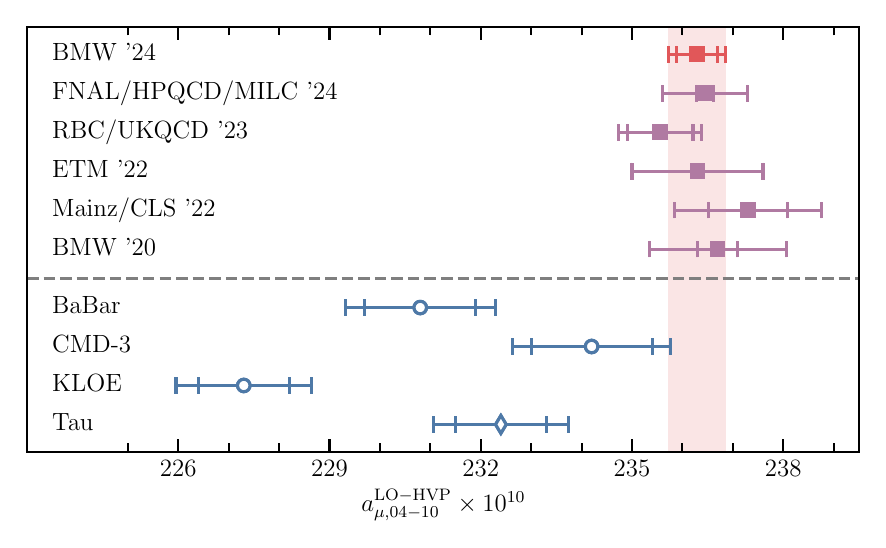}
    \caption
    {
	Comparison of our full intermediate-window results,
	$a_{\mu,04-10}^\text{LO-HVP}$, with others in the literature.
	In the top panel, we show lattice results~\cite{Boccaletti:2024guq,MILC:2024ryz,
	RBC:2023pvn,ExtendedTwistedMass:2022jpw,Ce:2022kxy,Borsanyi:2020mff}.
	In the lower panel, we show data-driven results~\cite{Davier:2023fpl} that use the
	measurements of the two-pion spectrum obtained in individual
	electron-positron annihilation experiments and in $\tau$-decays, as
	explained in \refcite{Davier:2023fpl}.
    }
    \label{fig:idwin}
\end{figure}

One window of particular interest focuses on an intermediate region of
Euclidean time between \qtylist{0.4;1.0}{fm}. This specific window is
particularly accessible to lattice QCD calculations, and has been computed
by many collaborations, with excellent agreement as shown in \fig{fig:idwin_light}.
This window emphasises energies near the rho resonance, which is the
region in which tension between data-driven inputs is greatest. This can
be seen from the spread between data-driven determinations based on different
experiments seen in \fig{fig:idwin}.

\begin{figure}[tbp!]
    \centering
    \includegraphics[scale=0.8]{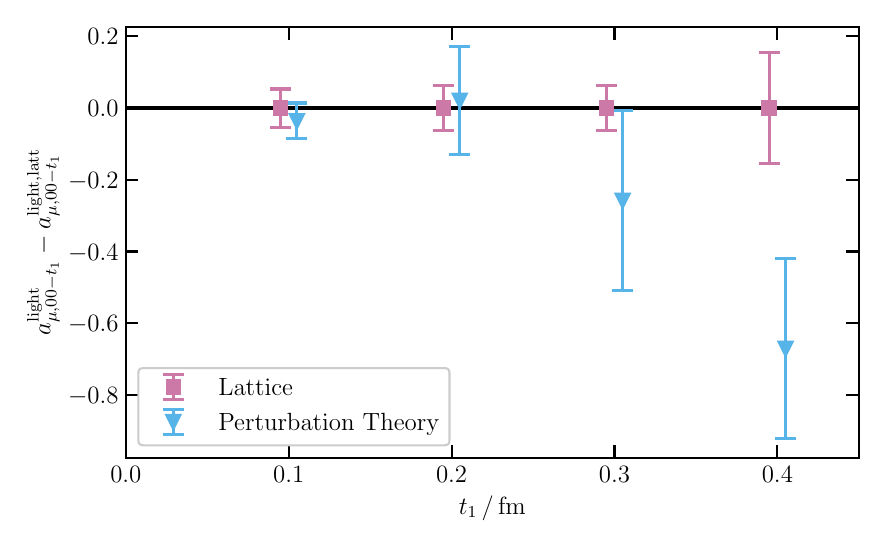}
    \caption
    {
	Comparison of the short-distance window observables
	$a_{\mu,00-t_1}^\mathrm{light}$ between the lattice
	results and those obtained from	perturbation theory.
	For ease of comparison, the central value of the lattice
	results has been subtracted off, and a slight
	horizontal offset has been applied.
    }
    \label{fig:sd_vs_pert}
\end{figure}

\begin{figure}[tbp!]
    \centering
    \includegraphics[scale=0.8]{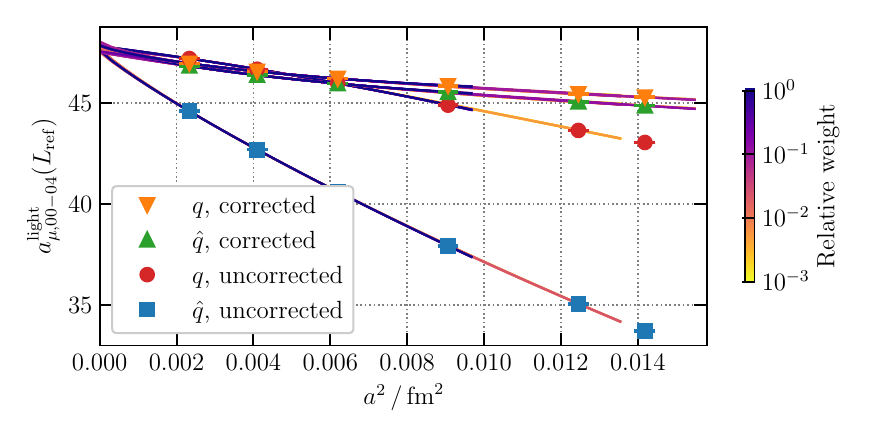}\\
    \includegraphics[scale=0.8]{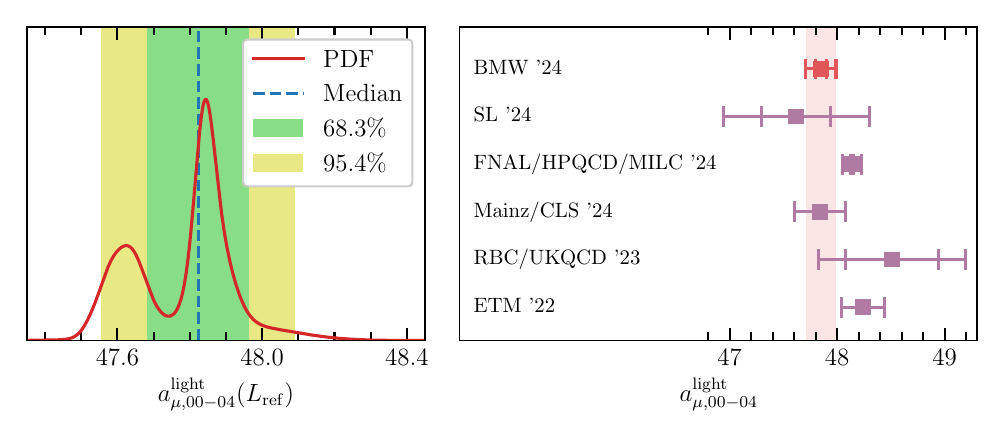}
    \caption
    {
	\label{fig:sdwin} Light-connected
	window observable $a_{\mu,00-04}^\mathrm{light}$.
	Top: continuum extrapolations, coloured according to the weight of
	the fit in our analysis.
	We plot fits with two different kernel
	functions, denoted by $q$ and $\hat{q}$, which correspond to two different
	discretisations for the momentum, and also show fits to data either corrected or not
	corrected by tree-level perturbation theory.
	Lower left: the probability distribution
	function, which displays two dominant peaks, corresponding to the variation
	between $q$ and $\hat{q}$ in the uncorrected case.
	Lower right: we compare our result with others from the
	literature~\cite{ExtendedTwistedMass:2022jpw,RBC:2023pvn,Kuberski:2024bcj,MILC:2024ryz,Spiegel:2024dec}.
    }
\end{figure}

The short distance part below \qty{0.4}{fm} exhibits important logarithmic
discretisation effects.
Perturbation theory agrees well with the lattice results below \qty{0.3}{fm},
as seen in \fig{fig:sd_vs_pert}.
It successfully quantifies these logarithmic effects, so we use a perturbation
theory motivated fit form to account for them.
The results for this window are shown in \fig{fig:sdwin}, including a
comparison to other lattice groups' calculations that shows good agreement.

\section{Hybrid Approach\label{sec:hybrid}}

The long-distance region is very sensitive to finite-size effects, and to
statistical uncertainties, particularly in the region beyond \qty{2.8}{fm}.
This makes it the most challenging part of the integral to compute in lattice QCD.
It only contributes about \(5\%\) of the final value, but it contributes
significantly to the final uncertainty.

\begin{figure}[tb]
    \centering
    \includegraphics[width=0.48\textwidth]{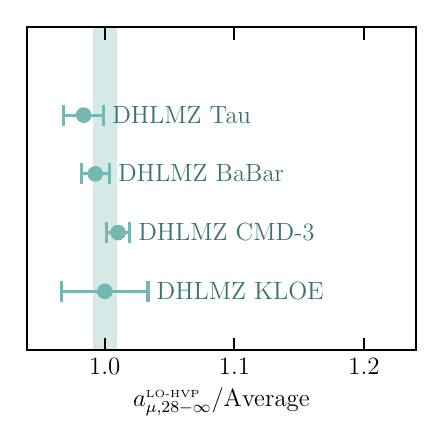}~%
    \includegraphics[width=0.48\textwidth]{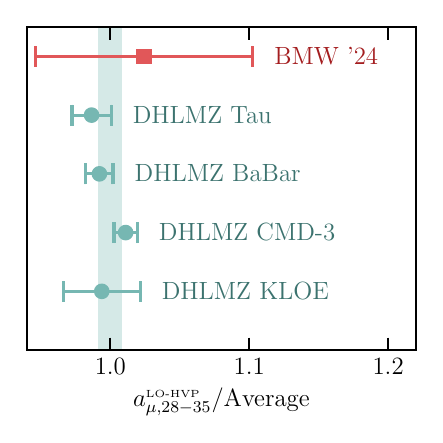}
    \caption
    {
	A comparison of data-driven results for the long-distance tail of the HVP with two-pion production
	data restricted to specific experimental data sets, including results from tau decays~\cite{Boccaletti:2024guq}.
	The shaded bands represent an average of the data-driven points, and all points are relative to the central
	value of that average.
	Left: the entire tail from \qty{2.8}{fm} to infinity.
	Right: a comparison with lattice for the part of the tail from \qty{2.8}{fm} to \qty{3.5}{fm}.
	The scatter of the individual results is consistent
	with their uncertainties, and the lattice result agrees well, although its errors are larger. The
	shaded band is a combination of the data-driven points.
    }
    \label{fig:tailcomp}
\end{figure}

However, it turns out that this very-long-distance tail is dominated by
low energy states, well below the rho resonance.
In particular, the experimental tensions that plague the data-driven approach
in the region of the rho peak are significantly suppressed by this kernel.
This makes this window one of the most reliable to compute in the data-driven
approach.
As shown in \fig{fig:tailcomp}, the data-driven results based on different
experiments are in good agreement with one another and also with the lattice
result.

This means we can take advantage of the complementary strengths of the lattice
calculation and data-driven determination by adopting a hybrid approach.
By replacing this small part of the integral with a data-driven result, the
finite-size uncertainty is significantly reduced, and the statistical and
physical point uncertainties are also improved.
At the same time, we avoid sensitivity to the data-driven discrepancies and
restrict the data-driven contribution to less than 5\% of the total.

To account for possible sensitivities to how the data-driven part of the hybrid
calculation is performed, we include additional uncertainties associated with
variations in the data-driven procedure, specifically:
\begin{itemize}
    \item whether the different experiments are averaged before or after integrating
	over the energy spectrum,
    \item whether additional experiments that only cover part of the energy range
	are included or not,
    \item the effect of including or excluding the CMD3 experiment, and
    \item the difference between the BaBar and KLOE experiments on the rho peak.
\end{itemize}
Including these additional systematics gives an overestimate of the final
uncertainties, as it will multiply count several sources of uncertainties.
It ends up increasing the uncertainty on the tail contribution by a factor of
two but has a negligible effect on the uncertainty of the final result for
\(a_\mu\).

\begin{figure}[tbp]
    \centering
    \includegraphics[scale=0.8]{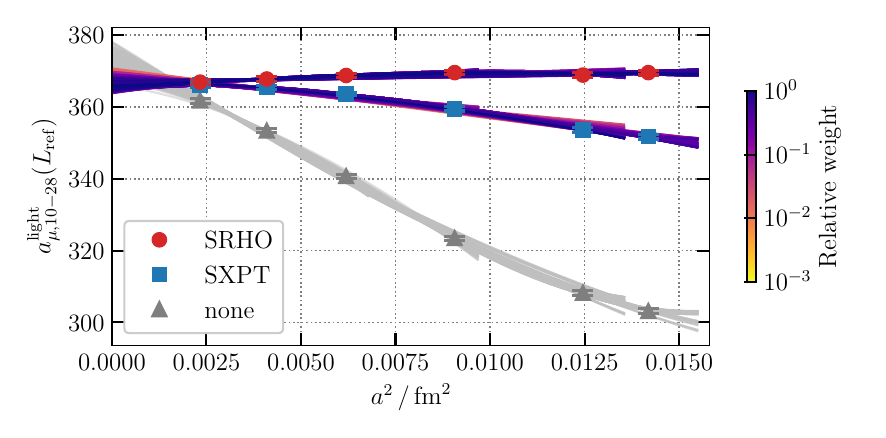}\\
    \includegraphics[scale=0.8]{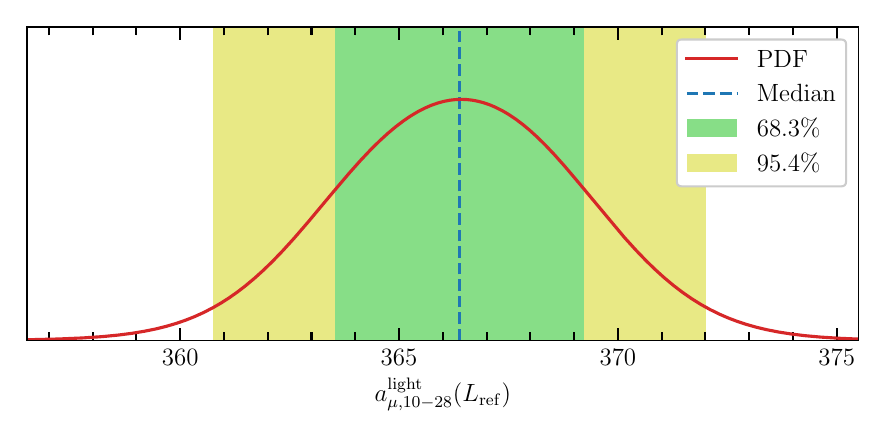}
    \caption
    {
	\label{fig:ldwin} Light-connected
	window observable $a_{\mu,10-28}^\mathrm{light}$.
	Top: continuum extrapolations, coloured according to the weight of
	the fit in our analysis, with no, NNLO XPT and SRHO taste
	improvements.
	Bottom: the probability distribution function.
    }
\end{figure}

The final Euclidean time region from \qty{1.0}{fm} to \qty{2.8}{fm} is then
computed in lattice QCD. The results are shown in \fig{fig:ldwin}.

\section{Scale Setting}

\renewcommand{\arraystretch}{1.0}
\begin{figure}[b!]
    \centering
    \includegraphics[scale=0.8]{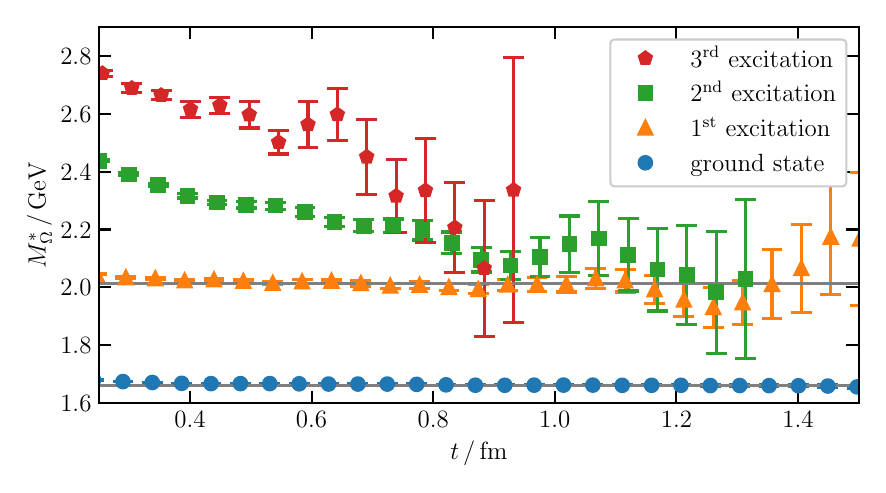}
    \caption
    {
	Ground state $\Omega$ baryon effective mass obtained through a GEVP,
	along with three excited states, at $\beta=4.1479$.
	The band through the ground state is from a single-exponential
	fit to the projected ground state propagator.
	The line through the first excitation shows the
	mass of the $\Omega$ excitation observed at Belle~\cite{Belle:2018mqs}.
    }
    \label{fig:omega} 
\end{figure}

The 2020 result used the mass of the Omega baryon to set the physical scale.
Our most precise result for the Omega mass uses a generalised eigenvalue
problem (GEVP) to isolate the ground state. We use a combination of spatial
Wuppertal smearing~\cite{Gusken:1989ad} and the Generalised-Pencil-of-Function
approach~\cite{Aubin:2010jc} to construct a \(6\times6\) correlation matrix.
The GEVP successfully extracts four states: the ground state Omega, and three
excited states that have energies consistent with the expected resonances,
based on experimental results~\cite{Belle:2018mqs} and quark model
predictions~\cite{Capstick:1986ter}.
This allows for a clean and clear ground state plateau as shown in
\fig{fig:omega}.
However, this GEVP does not isolate the nearby \(K\Xi\) scattering states,
and hence there are potential contaminations from these states that may
not be controlled.

\begin{figure}[b!]
    \centering
    \includegraphics[scale=0.8]{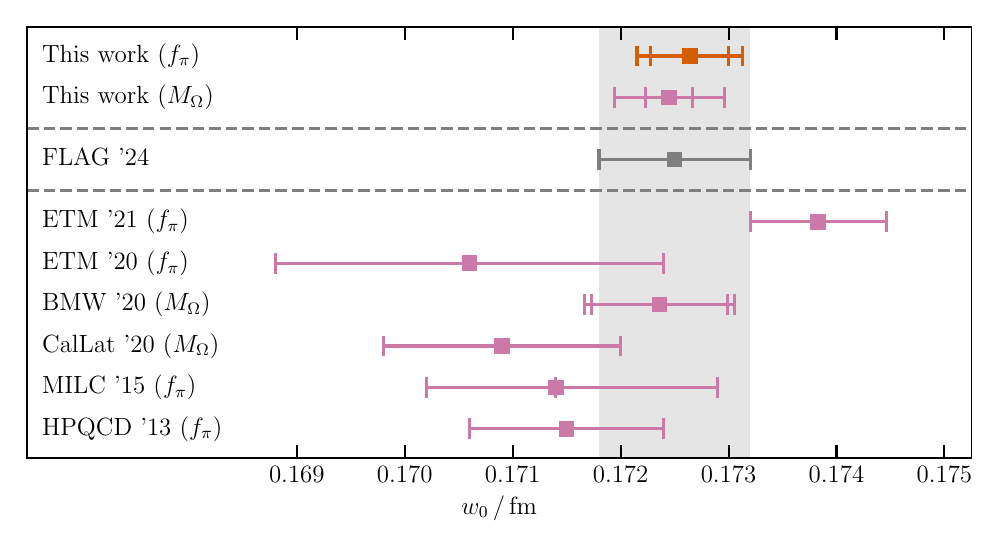}
    \caption
    {
	Comparison of the gradient-flow scale $w_0$ using either \(M_\Omega\)
	or \(f_\pi\) for scale setting.
	The lower panel shows earlier determinations~\cite{Borsanyi:2020mff,
        ExtendedTwistedMass:2021qui,Miller:2020evg,MILC:2015tqx,Dowdall:2013rya,
	ExtendedTwistedMass:2020tvp} and
	the grey band shows the latest FLAG
	average~\cite{FlavourLatticeAveragingGroupFLAG:2024oxs}.
    }
    \label{fig:scales}
\end{figure}

To avoid this potential unquantified systematic, we have since moved to a
scale setting based on the pion decay constant \(f_\pi\).
This required a determination of the radiative corrections to leptonic
decays, which were partly taken from the electro-quenched calculation of
\refcite{DiCarlo:2019thl}, and partly computed in our 2024 work.
This analysis was done completely blinded, and gave a
result that was completely consistent with our \(M_\Omega\) scale, as
seen in \fig{fig:scales}.

\section{Change From 2020}

It is useful to quantify the agreement or disagreement between our 2020 and
2024 results. The major changes that have occurred between the two are:
\begin{itemize}
    \item the addition of a new, finer lattice spacing,
    \item subdivision of the integral into individual windows,
    \item the inclusion of fits that are cubic in \(a^2\) (in 2020,
	there were only linear and quadratic),
    \item changing the quantity used for scale setting (\(M_\Omega\)
	to \(f_\pi\)), and
    \item the replacement of the tail with a data-driven result
	(significantly reducing finite-size effect).
\end{itemize}
With all of these factors taken together, we estimate the correlated
difference between the old and new results to be \(7.6(5.2)\times 10^{-10}\).
This means the new result is \(1.5\sigma\) higher. This is entirely
consistent with expected statistical fluctuations.

\section{Conclusion}

Recent lattice QCD results have surpassed the precision of all other theoretical predictions
of the hadronic vacuum polarisation contribution to the muon magnetic moment.
When taken together with the latest theory consensus for the other
contributions~\cite{Aliberti:2025beg}, these results show excellent agreement with the latest
experimental measurements~\cite{Muong-2:2025xyk}. This a remarkable success for quantum field
theory, bringing together diverse computational tools to include all aspects of the Standard
Model in a single calculation that validates the Standard Model to 0.31 ppm.

\section{Acknowledgements}
Figures in this work were produced with the aid of Matplotlib~\cite{Hunter:2007}.

We gratefully acknowledge GCS e.V., GENCI (grant 502275), EuroHPC Joint Undertaking (grants EXT-2023E02-063, EXT-2024E02-109) and NCMAS for providing computer time on SuperMUC-NG at
LRZ; HAWK and HUNTER at HLRS; JUWELS, JURECA and JUPITER at FZJ; Joliot-Curie/Irène Rome at TGCC;
Jean-Zay V100 at IDRIS; Adastra at CINES; Leonardo at CINECA; LUMI at CSC; Gadi at NCI; and Setonix at PSC.
We also gratefully acknowledge grants AMX-18-ACE-005, AMX-22-RE-AB-052, NW21-024-A, BMBF-05P21PXFCA, ERC-MUON-101054515, NW21-024-B, DOE-0000278885 and DE-SC0025025; and the Ramsay Fellowship.

\bibliographystyle{JHEPMod}
\bibliography{article}

\end{document}